\documentclass[twocolumn,english,aps,prl,reprint,nofootinbib,superscriptaddress]{revtex4-1}
\usepackage[T1]{fontenc}
\usepackage[latin9]{inputenc}
\usepackage{amsmath}
\usepackage{amssymb}
\usepackage{graphicx}

\usepackage{color}
\usepackage{float}
\usepackage{hyperref}

\hypersetup{
    colorlinks,
    linkcolor={black},
    citecolor={black},
    urlcolor={blue}
}

\makeatletter
\def\clearfmfn{\let\@FMN@list\@empty}    % <- clears the list of frontmatter footnotes
\@ifundefined{textcolor}{}
{%
 \definecolor{BLACK}{gray}{0}
 \definecolor{WHITE}{gray}{1}
 \definecolor{RED}{rgb}{1,0,0}
 \definecolor{GREEN}{rgb}{0,1,0}
 \definecolor{BLUE}{rgb}{0,0,1}
 \definecolor{CYAN}{cmyk}{1,0,0,0}
 \definecolor{MAGENTA}{cmyk}{0,1,0,0}
 \definecolor{YELLOW}{cmyk}{0,0,1,0}
}

\makeatother

\usepackage{pslatex}

\usepackage{babel}

\begin{document}

\title{Direct Laser Cooling of a Symmetric Top Molecule}

\author{Debayan Mitra}
\thanks{These authors contributed equally. dmitra@g.harvard.edu; \\ vilas@g.harvard.edu}
\affiliation{Harvard-MIT Center for Ultracold Atoms, Cambridge, MA 02138, USA}
\affiliation{Department of Physics, Harvard University, Cambridge, MA 02138, USA}

\author{Nathaniel B. Vilas}
\thanks{These authors contributed equally. dmitra@g.harvard.edu; \\ vilas@g.harvard.edu}
\affiliation{Harvard-MIT Center for Ultracold Atoms, Cambridge, MA 02138, USA}
\affiliation{Department of Physics, Harvard University, Cambridge, MA 02138, USA}

\author{Christian Hallas}
\affiliation{Harvard-MIT Center for Ultracold Atoms, Cambridge, MA 02138, USA}
\affiliation{Department of Physics, Harvard University, Cambridge, MA 02138, USA}

\author{Lo\"{i}c Anderegg}
\affiliation{Harvard-MIT Center for Ultracold Atoms, Cambridge, MA 02138, USA}
\affiliation{Department of Physics, Harvard University, Cambridge, MA 02138, USA}

\author{\\Benjamin L. Augenbraun}
\affiliation{Harvard-MIT Center for Ultracold Atoms, Cambridge, MA 02138, USA}
\affiliation{Department of Physics, Harvard University, Cambridge, MA 02138, USA}

\author{Louis Baum}
\affiliation{Harvard-MIT Center for Ultracold Atoms, Cambridge, MA 02138, USA}
\affiliation{Department of Physics, Harvard University, Cambridge, MA 02138, USA}

\author{Calder Miller}
\affiliation{Harvard-MIT Center for Ultracold Atoms, Cambridge, MA 02138, USA}
\affiliation{Department of Physics, Harvard University, Cambridge, MA 02138, USA}

\author{Shivam Raval}
\affiliation{Harvard-MIT Center for Ultracold Atoms, Cambridge, MA 02138, USA}
\affiliation{Department of Physics, Harvard University, Cambridge, MA 02138, USA}

\author{John M. Doyle}
\affiliation{Harvard-MIT Center for Ultracold Atoms, Cambridge, MA 02138, USA}
\affiliation{Department of Physics, Harvard University, Cambridge, MA 02138, USA}

\date{\today}

\begin{abstract}
We report direct laser cooling of a symmetric top molecule, reducing the transverse temperature of a beam of calcium monomethoxide (CaOCH$_3$) to $1.8\pm0.7$ mK while addressing two distinct nuclear spin isomers. These results open a path to efficient production of ultracold chiral molecules and conclusively demonstrate that by using proper rovibronic optical transitions, both photon cycling and laser cooling of complex molecules can be as efficient as for much simpler linear species. 
\end{abstract}
\maketitle

Laser cooling of atomic systems has enabled extraordinary progress in quantum simulation, precision clocks, and quantum many-body physics~\cite{phillips1998nobel,bloch2008many,bohn2017cold,safronovanewphysics2018}. Extending laser cooling to a diversity of complex polyatomic molecules would provide qualitatively new and improved platforms for these fields. The parity doublets that result from rotations of a molecule around its principal axis, a general feature of symmetric top molecules, give rise to highly polarized states with structural features that are greatly desirable for both quantum science and precision measurement~\cite{wall2015realizing,yu2019scalable, kozyryev2017precision}. The rapid photon cycling that is necessary for laser cooling additionally provides a key to many envisioned applications that benefit from high-fidelity quantum state readout. However, the same complexity that provides these advantages makes laser cooling challenging for these molecules. Recent theoretical proposals have nonetheless suggested that laser cooling of polyatomic molecules, even nonlinear ones like symmetric tops, is a practical possibility~\cite{Isaevproposal2016, kozyryev2016proposal,AugenbraunAsymmetric2020}.

In this work, we use rapid photon cycling to laser cool a molecular beam of calcium monomethoxide (CaOCH$_3$), reducing the transverse temperature from $22\pm1$~mK to $1.8\pm0.7$~mK while scattering over 100 photons. We demonstrate efficient and state-selective cooling of two nuclear spin isomers (NSIs)~\cite{Sun2005,Kravchuk319,Sun2015}. The laser cooling methods applied here result in rapid damping of molecular motion on a submillisecond timescale, without the need for trapping, and open a path to efficient production of ultracold nonlinear molecules, including their eventual use in precision measurements and optical tweezer arrays \cite{Anderegg1156,kozyryev2017precision}.

Laser cooling relies on repeatedly scattering photons from an atom or molecule via rapid optical cycling. This removes energy and entropy with directed momentum kicks and spontaneous emission events. Both direct~\cite{Lemeshko2013manipulation, fulton2004optical,lavert2011moving,vanhaecke2007multistage,zeppenfeld2012sisyphus,prehn2016optoelectrical,wu2017cryofuge,liu2017magnetic} and indirect~\cite{Ni231,de2019degenerate} methods of slowing, cooling, and trapping molecules have been employed with considerable success. Direct laser cooling has brought diatomic~\cite{shuman2010laser,truppe2017molecules,anderegg2017radio,collopy20183d} and linear triatomic~\cite{kozyryev2017sisyphus,augenbraun2020laser,Baum2020} molecules into the microkelvin regime, with orders of magnitude increase in phase-space density. Critically, the ability to rapidly cycle photons, which is essential to laser cooling, naturally also allows for efficient quantum state preparation and readout \cite{cheuklambda2018,Anderegg1156}, necessities for proposed quantum computation and simulation platforms using ultracold molecules, including those proposed for symmetric top molecules~\cite{wall2015realizing,yu2019scalable}.

\begin{figure}[ht]
\begin{centering}
\includegraphics[width = 0.5\textwidth]{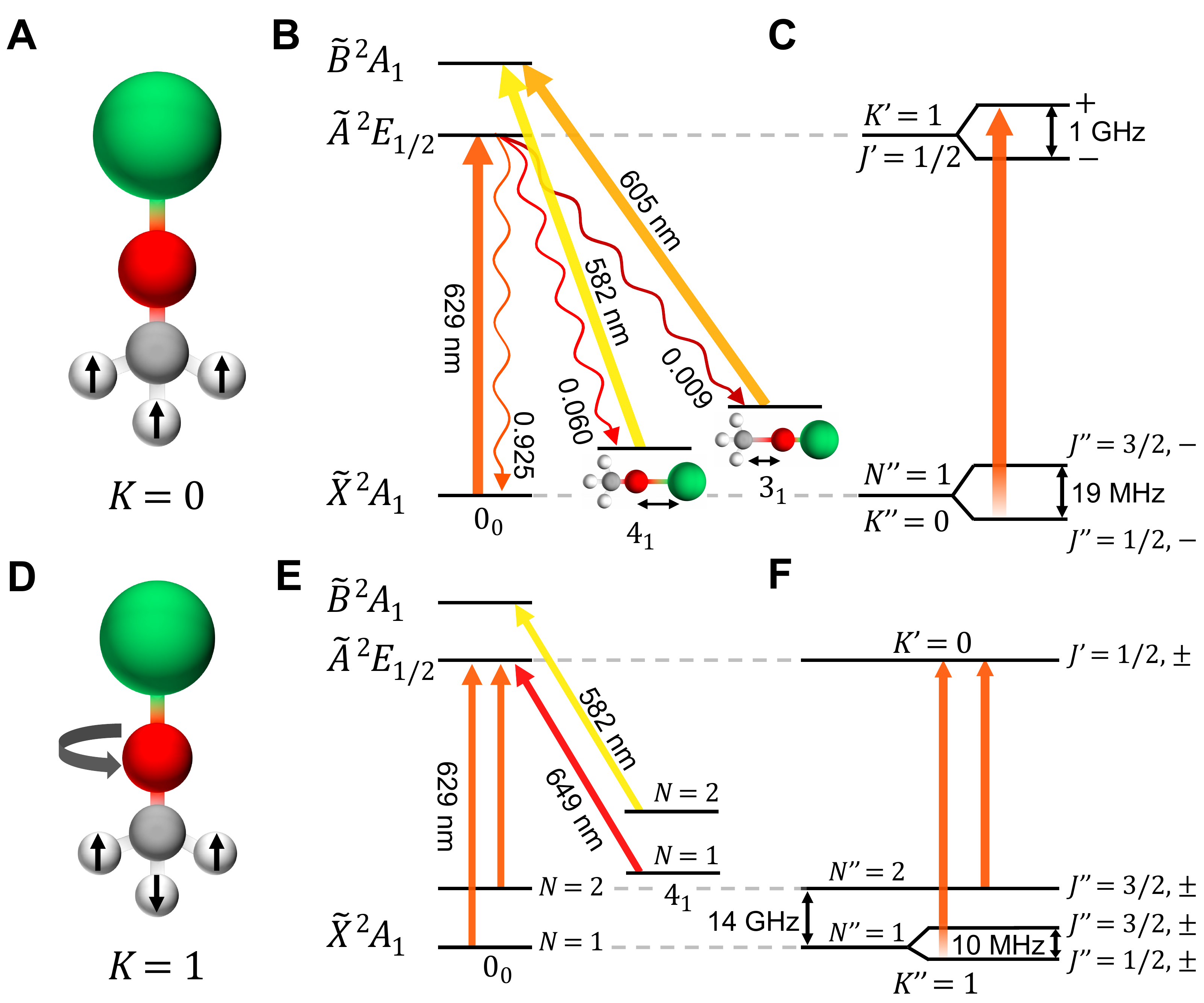}
\caption{\label{fig:structure} \textbf{CaOCH$_3$ laser cooling schemes.} (\textbf{A} and\textbf{D}) We employ two optical cycling schemes, which target molecules differing in their angular momentum quantum number $K$ and their nuclear spin statistics (represented by arrows on the hydrogen nuclei). (\textbf{B} and \textbf{C}) For ortho-CaOCH$_3$, we address the $0_0$, $4_1$, and $3_1$ vibrational levels of the $\tilde{X}\,^2A_1(K''=0)$ ground state. Vibrational branching ratios are illustrated by downward arrows. Rotational closure is achieved by addressing $N''=1$, $J''=1/2,3/2$ ground-state manifolds. The total parity of each state is indicated by $+$ and $-$ signs. (\textbf{E} and \textbf{F}) For para-CaOCH$_3$, we address the $0_0$ and $4_1$ vibrational levels of the $\tilde{X}\,^2A_1(K''=1)$ electronic ground state, driving transitions from $N''=1$ and $N''=2$ to achieve rotational closure. Each $J$ state contains an unresolved parity doublet denoted by $\pm$. See Supplemental Materials for further details on the optical cycling scheme.}
\par\end{centering}
\end{figure}

The established recipe for achieving optical cycling and laser cooling of molecules requires three key ingredients: strong electronic transitions between two fully bound molecular states; diagonal Franck-Condon factors (FCFs), which limit branching to excited vibrational levels; and rotationally closed transitions. Here we satisfy these conditions for the molecule CaOCH$_3$ using two distinct optical cycling schemes that enable rapid scattering of photons. Efficient laser cooling is demonstrated using only a few lasers, despite the presence of twelve vibrational modes. The state-selectivity of the cooling is intimately connected to both the nuclear spin statistics of the molecule, as well as its rigid-body angular momentum along the symmetry axis, denoted by the quantum number $K''$ in the ground state (Fig. \ref{fig:structure}, A and D). These distinctive features of symmetric top molecules were not accessible to previously laser-cooled diatomic and triatomic molecules.

We study laser cooling of both nuclear-spin isomers (NSIs) of CaOCH$_3$, each of which corresponds to a specific set of $K$ states. To cool the symmetric (ortho) NSI, we laser excite molecules in ground states with $K''=0$ (Fig. \ref{fig:structure}A-C). The main cooling laser light at 629 nm drives the diagonal $\tilde{X}\,^2A_1 - \tilde{A}\,^2E_{1/2} 0_0^0$ vibronic transition between states with no vibrational excitations.\footnote{Here we use the standard vibrational notation $n_{v''}^{v'}$, where $n$ labels the normal mode of vibration, and $v''$ and $v'$ specify the number of excited quanta in the lower ($v''$) and upper ($v'$) state. We adopt the labeling convention where $n=3$ is the antisymmetric (O--C) stretching mode, $n=4$ is the symmetric (Ca--O) stretching mode, and $n=8$ is the doubly degenerate Ca--O--C bending mode (see Supplemental Materials). The remaining eight modes play no role in this work. Individual ground and excited states are labeled $n_{v''}$ and $n^{v'}$, respectively.} Parity and angular momentum selection rules allow full rotational closure addressing a single rotational component ($N''=1$) of the $\tilde{X}\,^2A_1$ ground state (Fig. \ref{fig:structure}C). This is the same scheme employed for laser cooling of linear molecules, indicating that symmetric top states with $K''=0$ effectively "freeze out" the particular additional complexity of nonlinear CaOCH$_3$ molecules for this specific NSI. We use two lasers to take the molecules that are lost to the $4_1$ and $3_1$ vibrational manifolds and repump them back into the cooling cycle, as depicted in Fig. \ref{fig:structure}B. The diagonal FCFs of CaOCH$_3$ \cite{kozyryev2019determination} enable each molecule to scatter an average of 120 photons before being lost to other vibrational states. These losses are understood to be dominated by decay to the $4_2$ and $3_1 4_1 8_1$ vibrational levels of the $\tilde{X}\,^2A_1$ state \cite{kozyryev2019determination}.

We also laser cool the asymmetric (para) isomer, exciting ground states that have $K''=1$ (Fig. \ref{fig:structure}D-F). The existence of unresolved (opposite) parity doublets in both the ground and excited states means that full rotational closure requires addressing both $N''=1$ and $N''=2$ components of the $\tilde{X}\,^2 A_1$ state. The main cooling laser at 629 nm addresses the $\tilde{X}\,^2A_1(N''=1) - \tilde{A}\,^2E_{1/2}0^0_0$ vibronic transition, while rotational and vibrational repumping lasers recover molecules from the $\tilde{X}\,^2A_1(N''=2)0_0$ and $\tilde{X}\,^2A_1(N''=1)4_1$ states (Fig. \ref{fig:structure}E). This enables scattering an average of 30 photons before molecules are optically pumped into the $\tilde{X}\,^2A_1(N''=2)4_1$ state. We recover this population by optical pumping via the $\tilde{B}\,^2A_10^0$ state before detecting the molecules. We note that if one were to address both the $N''=1$ and $N''=2$ components of the $3_1$ vibrational mode, this would allow scattering 120 photons on average, just as in the ortho-CaOCH$_3$ scheme described earlier. See Supplemental Materials for further details on both the para and ortho photon cycling approaches.

We laser cool CaOCH$_3$ by employing the magnetically-assisted Sisyphus effect, a highly efficient and robust cooling method first established with atoms~\cite{emile1993,sheehymagnetic1990,shuman2010laser,kozyryev2017sisyphus,lim2018laser,augenbraun2020laser}. Molecules are produced in a buffer gas beam source \cite{hutzler2012buffer}, forming a molecular beam that passes through a standing wave of near-resonant light containing all of the optical frequencies necessary to establish optical cycling. The molecules in ground state sublevels that couple to the cooling light (``bright states'') move through the periodic, AC Stark-shifted potential that is created by the standing wave. As a molecule approaches an antinode it is optically pumped into a dark sublevel with very small AC Stark shift. A small magnetic field of magnitude $|\vec{B}| \approx 1$~G, aligned at an angle $\theta = 45^\circ$ relative to the laser polarization axis, remixes dark and bright sublevels preferentially at the nodes, restarting the process. When the primary laser frequency has a positive detuning ($\Delta > 0$) with respect to the main cooling transition ($\tilde{X}\,^2A_1 - \tilde{A}\,^2E_{1/2} 0_0^0$), antinodes of the standing wave correspond to peaks of the AC-Stark shifted potential, and molecules lose energy as they climb the potential hill before being pumped to a dark state, leading to cooling. The opposite process occurs for $\Delta < 0$, resulting in Sisyphus heating.

\begin{figure*}[ht]
\begin{centering}
\includegraphics[width = 1\textwidth]{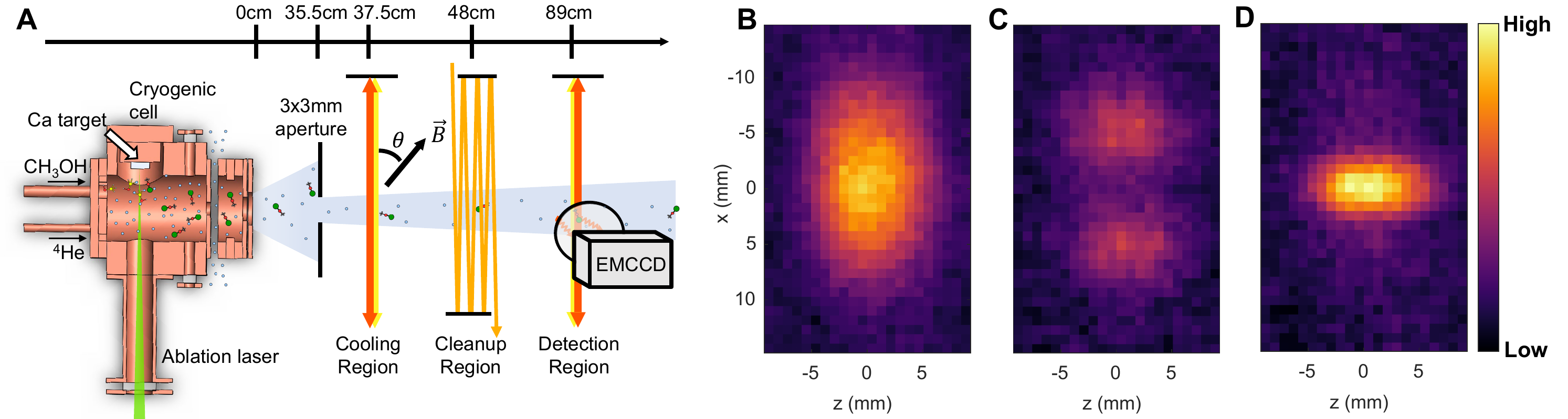}
\caption{\label{fig:apparatus} \textbf{Apparatus and beam images.} \textbf{(A)} Schematic of the experimental apparatus, illustrating the beam source, laser cooling, cleanup, and detection regions (not to scale). The cooling region contains a near-resonant standing wave generated by retroreflecting a single, linearly polarized, 6~mm $1/e^2$ diameter Gaussian laser beam. In the cleanup region, molecules are repumped out of the $\tilde{X}\,^2A_1 3_1$ (ortho-CaOCH$_3$) and $\tilde{X}\,^2A_1(N''=2)4_1$ (para-CaOCH$_3$) states before being imaged onto an EMCCD camera via laser-induced fluorescence (LIF) detection. Beam images for ortho-CaOCH$_3$ ($K''=0$) are shown for \textbf{(B)} unperturbed, \textbf{(C)} Sisyphus heated ($\Delta = -15$~MHz), and \textbf{(D)} Sisyphus cooled ($\Delta = +25$~MHz) configurations.}
\par\end{centering}
\end{figure*}

The experimental apparatus is similar to one described previously \cite{Baum2020}. A schematic is shown in Fig. \ref{fig:apparatus}A. Briefly, CaOCH$_3$ molecules are produced in a cryogenic buffer gas environment by ablation of a calcium metal target in the presence of methanol vapor. The resulting beam has a mean forward velocity of 150$\pm$30 m/s and is collimated to a transverse temperature of $\sim 22$~mK by a $3\times 3$~mm square aperture immediately preceding the cooling region. After laser cooling, the molecules propagate $\sim50$ cm and undergo time-of-flight expansion in the direction transverse to propagation, mapping the momentum distribution onto the spatial profile of the beam. During this expansion they interact with vibrational repumping laser light that returns them to states that can be detected. Finally, in the detection region, the molecules are addressed with resonant laser light and the resulting fluorescence is imaged onto an electron multiplying charge-coupled device (EMCCD) camera to extract spatial information, and thus their transverse temperature.

Figures \ref{fig:apparatus}B-D show representative beam images of the ortho NSI ($K''=0$) for unperturbed, Sisyphus heated ($\Delta < 0$), and Sisyphus cooled ($\Delta > 0$) configurations. The cooled beam exhibits clear compression with respect to the unperturbed beam along the cooling axis, indicating a reduced transverse velocity spread, while the heated beam is separated into two lobes symmetrically displaced from the center of the beam. Both images demonstrate strong optical forces manipulating the molecular velocity distribution.

\begin{figure}
\begin{centering}
\includegraphics[width = 0.85\columnwidth]{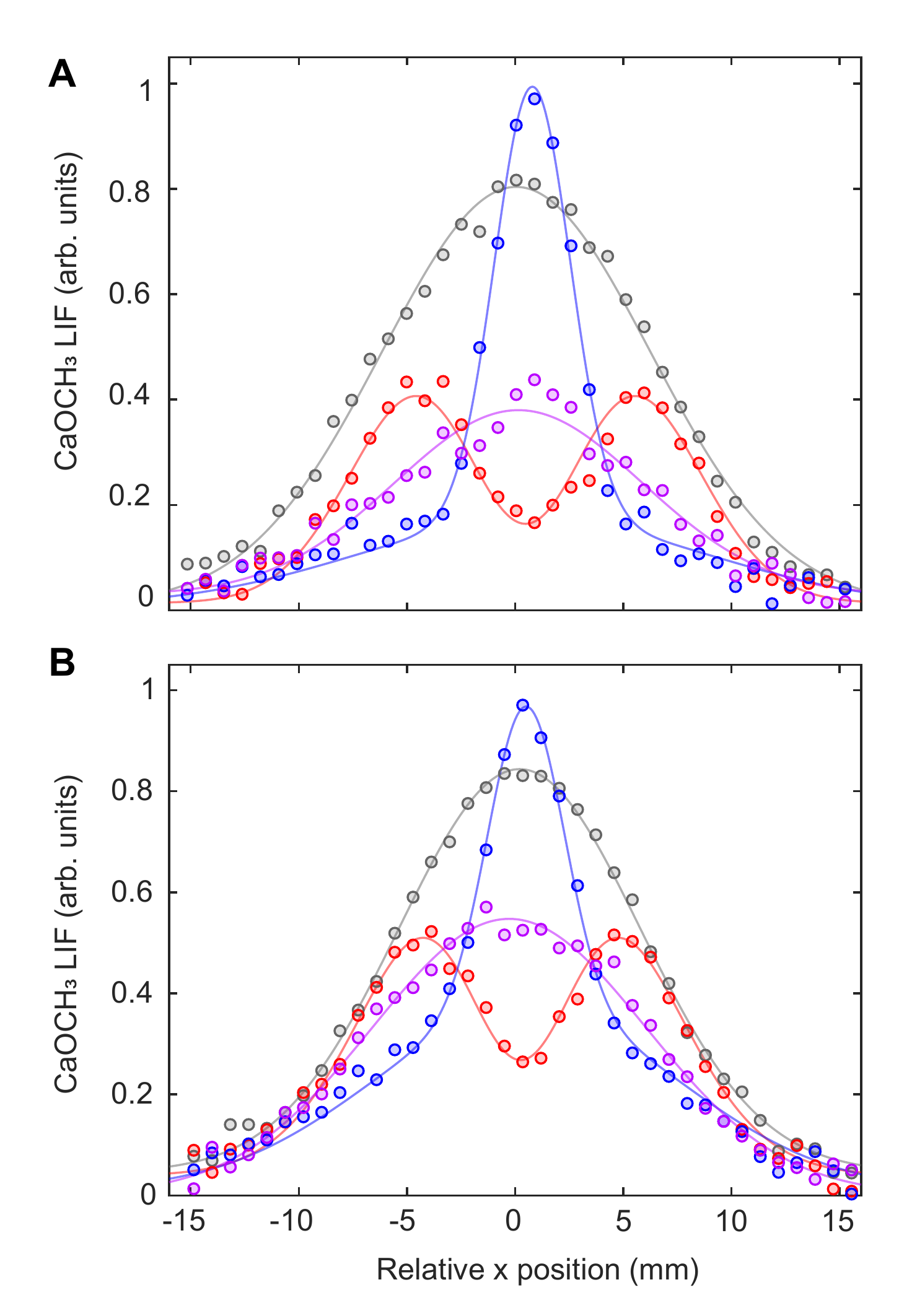}
\caption{\label{fig:beamtraces} \textbf{Sisyphus cooled and heated beam profiles.} Integrated laser induced fluorescence (LIF) vs. position for \textbf{(A)} ortho-CaOCH$_3$ ($K''=0$) and \textbf{(B)} para-CaOCH$_3$ ($K''=1$) cooling schemes. Sisyphus cooling at a positive detuning $\Delta$ = $+25$~MHz ($\Delta$ = $+20$~MHz) for the ortho (para) isomer manifests itself as a narrowing of the detected distribution (blue), while Sisyphus heating appears as a bimodal distribution ($\Delta$ = $-15$~MHz for both; red). Unperturbed (cooling lasers off; gray) and resonantly depleted ($\Delta$ = 0; purple) profiles have the same width but different integrated area due to optical pumping into dark vibrational states. Solid curves are Gaussian fits as described in the SM.}
\par\end{centering}
\end{figure}

By integrating along the direction of molecular beam propagation we obtain one-dimensional (1D) profiles, shown in Fig. \ref{fig:beamtraces} for both the ortho (A) and para (B) NSI cooling schemes. The cooled and heated profiles fit well to a sum of Gaussian distributions with two distinct widths, corresponding to two classes of molecules, those that were Sisyphus laser cooled and those that were not. The cooled molecules are those with transverse velocities less than the capture velocity, $v<v_c$. Molecules with $v>v_c$ are instead subject predominantly to Doppler cooling and heating, depending on laser detuning. When $\Delta>0$, a large fraction of molecules fall within $v_c$ and are cooled into a central, narrow peak on top of a broad Doppler heated background. In the red-detuned case ($\Delta<0$), molecules slower than $v_c$ are heated while faster ones are Doppler cooled. This leads to a concentration of molecules at velocities where the Doppler and Sisyphus forces balance, corresponding to approximately $v_c$. From the positions of these two peaks we estimate a capture velocity $v_c \approx 1.5$ m/s for $K''=0$ cooling of the ortho NSI. Unperturbed and resonantly depleted profiles are shown in gray and purple, and fit well to single Gaussian profiles.
 
\begin{figure*}
\begin{centering}
\includegraphics[width = 2\columnwidth]{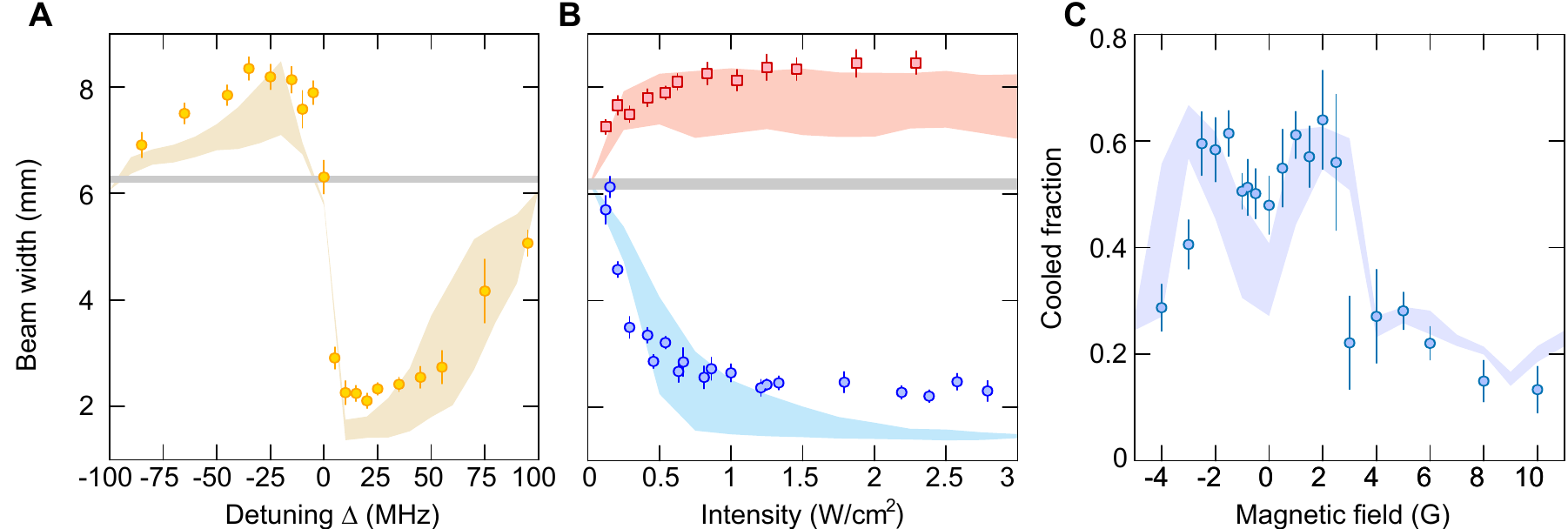}
\caption{\label{fig:analysis} \textbf{Experimental parameter scans and theory comparison.} (\textbf{A}) Variation of beam width as a function of detuning $\Delta$ at a cooling laser intensity of 400~mW/cm$^2$ and a magnetic field of $|B|$ = 1~G aligned at $45^\circ$ to the laser polarization axis. The grey band depicts the unperturbed molecular beam width. (\textbf{B}) Variation of beam width as a function of laser intensity for $\Delta = +25$~MHz (cooling; blue circles) and $\Delta = -15$~MHz (heating; red squares). A magnetic field $|B| = 1$~G was aligned at 45$^\circ$ to the laser polarization axis for the cooled molecules, while for the heated molecules a field $|B| = 2$~G was aligned at 90$^\circ$.  The observed beam width was only weakly dependent on this angle. (\textbf{C}) Variation of the fraction of cooled molcules as a function of magnetic field, taken at a laser intensity of 400~mW/cm$^2$, a detuning of $\Delta=+20$~MHz, and the magnetic field aligned at $90^\circ$ to the laser polarization axis. Shaded regions represent the results of the model described in the text, with error bands set by experimental uncertainty in the number of scattered photons. All error bars and bands for experimental data are standard error of the mean.}
\par\end{centering}
\end{figure*}

The integrated area of each of the three ortho-CaOCH$_3$ profiles with 1.1~W/cm$^2$ of cooling light applied is approximately 50\% that of the unperturbed profile (Fig \ref{fig:beamtraces}A). This effect is understood to be due to losses to vibrational states that are not repumped, most notably $\tilde{X}\,^2A_14_2$ and $\tilde{X}\,^2A_13_1 4_1 8_1$. Combining the observed depletion with branching ratios previously measured for CaOCH$_3$ \cite{kozyryev2019determination}, we determine that $80^{+100}_{-30}$ photons were scattered in the cooling process and $110^{+150}_{-40}$ photons were scattered in resonant depletion (see SM). From this observation we infer an average scattering rate of $\sim 2 \times10^6$ s$^{-1}$ across the cooling region, which is similar to scattering rates observed for laser cooling of diatomic and linear triatomic molecules \cite{barry2014magneto, anderegg2017radio, truppe2017molecules, collopy20183d, kozyryev2017sisyphus, augenbraun2020laser}. Finally, we determine the temperature of the molecules by fitting a Monte Carlo simulation of the molecular beam propagation to our data (see SM). This gives an initial transverse temperature $T_\perp = 22\pm1$~mK, which is reduced by Sisyphus laser cooling to $T_\perp = 1.8\pm0.7$ mK. Combined with the enhancement in on-axis molecule density seen in Fig \ref{fig:beamtraces}A, this 10x temperature reduction corresponds to a 4x increase in the on-axis phase-space density of the molecular cloud.\footnote{Here we define the phase-space density as $\rho = n \lambda^{dB}_x \lambda^{dB}_y \lambda^{dB}_z$, where $n$ is the number density and $\lambda^{dB}_i \sim 1/\sqrt{T_i}$ is the de Broglie wavelength defined in terms of the temperature along the $i$th spatial direction. Because we cool in 1D only, $\rho$ scales as $1/\sqrt{T}$.}

Figure \ref{fig:beamtraces}B shows beam profiles for para-CaOCH$_3$ ($K''=1$) laser cooling, taken at a laser intensity of 250 mW/cm$^2$. We observe significant Sisyphus cooling and heating, though the effect is weaker than for ortho-CaOCH$_3$ ($K''=0$). This is because more ground states are coupled to the same excited electronic state compared to the $K''=0$ scheme, leading to a lower scattering rate \cite{norrgard2016submillikelvin}. Additionally, the reduced laser intensity used, set by technical limitations, results in a smaller capture velocity and cooled fraction. Using the same analysis as above, we find that the cooled molecules here scatter an average of $25 \pm 10$ photons, corresponding to an estimated scattering rate of $\sim 0.75 \times 10^6$~s$^{-1}$. This is in qualitative agreement with an expected $\sim2$x reduction in scattering rate compared to ortho-CaOCH$_3$ due to the increased number of ground states (see SM). Cooling of the para NSI would be improved with higher laser intensity and/or interaction length.

We further characterize the Sisyphus cooling mechanism for ortho-CaOCH$_3$ by varying the laser detuning, intensity, and magnetic field as shown in Fig. \ref{fig:analysis}. The dependence of the molecular beam width, defined as the outermost radius at which the signal falls to $1/\sqrt{e}$ of its maximum value (see SM), on detuning ($\Delta$) is shown in Fig. \ref{fig:analysis}A. The sign of this dependence is opposite that of Doppler cooling and is a clear signature of the magnetically-assisted Sisyphus effect. Both Sisyphus heating and cooling are optimized at detunings of $\Delta\approx\pm20$~MHz, and we observe a $\sim 3$x decrease in molecular beam width for optimal cooling, corresponding to a significant increase in on-axis beam brightness. The dependence of the molecular beam width on cooling laser intensity is shown in Fig. \ref{fig:analysis}B. For both Sisyphus cooling ($\Delta>0$; circles) and Sisyphus heating ($\Delta<0$; squares), the beam width changes monotonically from its unperturbed value until saturating around an intensity of 600~mW/cm$^2$. This occurs at a maximum cooled fraction (defined as the portion of molecules captured by the Sisyphus effect and cooled into the narrow central peak) of 72(7)$\%$. This saturated value indicates that we are able to cool all molecules in our cryogenic beam with velocity $v_\text{trans} < v_c$, given the initial Gaussian velocity spread $\sigma_v \approx 1.6$ m/s $\approx v_c$. Finally, we plot the cooled fraction as a function of applied magnetic field for $\Delta>0$ (Fig. \ref{fig:analysis}C). As expected, Sisyphus cooling is suppressed at $|B|=0$, though incomplete cancellation of Earth's magnetic field nonetheless allows residual remixing of dark states. The cooling efficiency improves up to fields of $|B|\approx1-2$~G, which is consistent with the optimal field $B_0 \approx 1.6$~G predicted by equating the Larmor precession time due to the remixing field with the time for a molecule to travel from an antinode to a node of the standing wave (see SM). At large fields $|B|>3$~G we see a significantly weaker Sisyphus effect due to remixing of bright and dark states away from nodes of the standing wave.

We model ortho-CaOCH$_3$ laser cooling by solving optical Bloch equations for the density matrix of a molecule moving through the laser cooling region. Further details of this method may be found in Refs. \cite{emile1993, Devlin2016, Devlin2018}. Because power broadening is significant compared to the 19 MHz separation between the $J''=3/2$ and $J''=1/2$ ground states, we compute force profiles approximating the system as a single $J'' = 3/2 \leftrightarrow J' = 1/2$ transition, which captures the essential physics of Sisyphus laser cooling due to magnetic dark states in the system. Additional effects due to velocity-selective dark states could occur when all substructure is taken into account \cite{caldwell2019deep}, but they are not found to be significant here. The resulting computed force profiles are used in a Monte Carlo simulation to determine final molecular beam widths and cooled fractions (see SM). The results of the model are shown as shaded regions in Fig. \ref{fig:analysis}. We find that the damping coefficient at optimal detuning and magnetic field is $\eta_\text{max} \approx 3 \times 10^5 \text{ s}^{-1}$, which is several orders of magnitude higher than those achieved for standard radiative cooling of linear molecules, but expected for the Sisyphus mechanism \cite{truppe2017molecules,anderegg2017radio,Baum2020, emile1993}. From the force profiles, we also infer a capture velocity $v_c \sim 1.9$ m/s, which is consistent with the experimental estimate obtained from the beam profiles for the case of  $\Delta < 0$, as shown above.

Our results demonstrate the feasibility of directly laser cooling complex, nonlinear polyatomic molecules into the millikelvin temperature regime \cite{Ivanov2020,klos2019prospects}. This work opens the door to a number of future experiments that span a range of modern physical and chemical research frontiers. Because Sisyphus cooling is effective down to the recoil limit~\cite{hoogerland1992magnetically} (corresponding to $\sim 500$~nK for molecules with similar mass to CaOCH$_3$), these techniques could be used to achieve bright, highly-collimated, few-$\mu$K molecular beams useful for precision measurements and studies of ultracold chemistry \cite{lavert2014observation}. Efficient and state-selective laser cooling of both nuclear spin isomers also offers a method to separate them using radiation pressure beam deflection of specific spin species, a topic of interest in physical chemistry~\cite{kuppermanipulating2009,Sun2005,Kravchuk319,Sun2015}. By adding a small number of other laser frequencies to the laser cooling of CaOCH$_3$ \cite{kozyryev2019determination}, optical tweezer arrays of symmetric top molecules should be possible, as recently accomplished with diatomic species \cite{Anderegg1156}. These arrays would offer an ideal starting point for realizing new polyatomic quantum simulation and computation platforms \cite{wall2015realizing,yu2019scalable}. Laser cooling could also be extended to asymmetric tops, including  biochemically relevant chiral molecules~\cite{AugenbraunAsymmetric2020,kozyryev2016proposal,quack2002important,blackmondorigin2011}. Finally, laser cooling and trapping of the heavier symmetric top molecule YbOCH$_3$ would allow precise searches for time-reversal violating interactions at a previously inaccessible energy scale, while ultracold chiral molecules such as YbOCHDT could enable precision probes of fundamental parity violation \cite{kozyryev2017precision, augenbraun2020laser}.

This work was supported by the NSF, AFOSR, and ARO. We gratefully acknowledge Ivan Kozyryev for insightful discussions and comments on the manuscript, as well as Zack Lasner and Phelan Yu for useful discussions at various stages of this work. NBV acknowledges support from the NDSEG fellowship, LA from the HQI, and BLA from the NSF GRFP.

\bibliographystyle{apsrev4-1}
%\bibliography{CaOCH3Sisyphus_References}

%

\clearfmfn  

\pagebreak

\clearpage

\setcounter{equation}{0}

\setcounter{figure}{0}

\renewcommand{\thefigure}{S\arabic{figure}}

\renewcommand{\theequation}{S\arabic{equation}}

\noindent \Large\textbf{Supplemental Materials}
\bigskip

\begin{figure}
\begin{centering}
\includegraphics[width = 0.8\columnwidth]{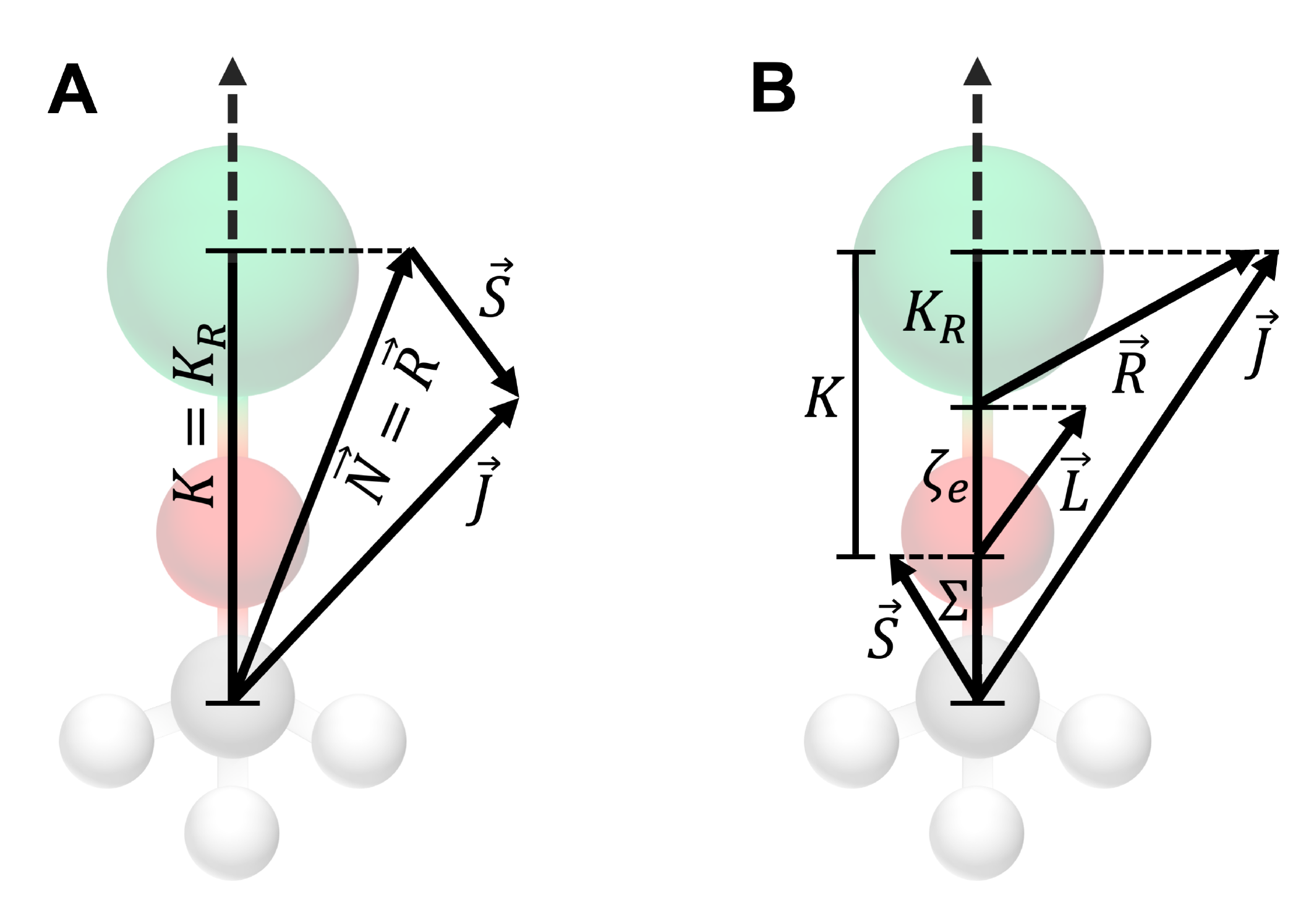}
\caption{\label{fig:hundscases} Quantum numbers for the electronic states of CaOCH$_3$ used in this work. The vertical axis is the (molecule-fixed) symmetry axis. \textbf{(A)} The $\tilde{X}\,^2A_1$ and $\tilde{B}\,^2A_1$ states are well described by Hund's case (b) with electronic orbital angular momentum $\vec{L} = 0$. $\vec{R}$ is the total rigid-body rotation and $K_R$ is its projection onto the symmetry axis. $\vec{N} = \vec{R} + \vec{L}$ also includes the electronic orbital angular momentum, and $K$ is its projection. $\vec{S}$ and $\vec{J}$ are the electron spin and total (excluding nuclear spin) angular momenta, respectively. \textbf{(B)} The $\tilde{A}\,^2E$ state is well described by Hund's case (a). This state has an electronic orbital angular momentum projection $\zeta_e \approx 1$, so that $K \approx K_R \pm 1$.}
\par\end{centering}
\end{figure}

\normalsize
\noindent\textbf{Optical cycling and CaOCH$_3$ structure}
\newline
We optically cycle photons on the $\tilde{X}\,^2A_1 - \tilde{A}\,^2E_{1/2}0^0_0$ transition in CaOCH$_3$. To prevent rotational branching we exploit angular momentum selection rules involving $K_R$, the projection of the total rigid body rotation $\vec{R}$ onto the molecular symmetry axis, and $K$, the projection of the total angular momentum excluding spin, $\vec{N}$. Note that the quantum number $K$ also includes the electronic angular momentum $\zeta_e$ about the symmetry axis. The relevant quantum numbers are illustrated in Fig. \ref{fig:hundscases}. In the non-degenerate states $\tilde{X}\,^2A_1$ and $\tilde{B}\,^2A_1$ we have $K \approx K_R$ while for the doubly-degenerate $\tilde{A}\,^2E$ state $|K| \approx  |K_R| \pm 1$. These two rotational levels are split by a large Coriolis interaction. Because almost all of the change in angular momentum comes from the electron's orbital angular momentum, $K_R$ is approximately conserved during excitation on the $\tilde{A}-\tilde{X}$ band. We therefore group states by $K_R$ and look for closed rotational transitions within these ``$K_R$ stacks'' \cite{brazierAX1989,marrmolecular1996}.

Within the $K_R = 0$ stack, we drive transitions from $\tilde{X}\,^2A_1(K''=0)$ to $\tilde{A}\,^2E_{1/2}(K'=1)$. The ground state is well described by Hund's case (b), with individual rotational levels $N''$ split into two components $J''=N''\pm1/2$ by the spin-rotation interaction, while the $\tilde{A}\,^2E$ state is well described by Hund's case (a). Rotational closure is attained by addressing $^rP_{11}$ and $^rQ_{12}$ transitions from $N''=1$, i.e. $\tilde{X}\,^2A_1(K''=0,N''=1,J''=1/2,3/2,-) \rightarrow \tilde{A}\,^2E_{1/2}(K'=1,J'=1/2,+)$, where signs indicate the parity of each state (see Fig. 1 of the main text).\footnote{We adopt the notation used in Ref. \cite{brown1971rotational} and label transitions $^{\Delta K}\Delta J_{F_1F_2}$, where $\Delta K = K'-K''$, $\Delta J = J'-J''$, and $F_1$ and $F_2$ denote the component character of the excited and ground state, respectively. $F_1 = 1$ applies to the $\Omega = 1/2$ spin-orbit component of the excited state, while $F_2 = 1,2$ correspond to the upper and lower spin-rotation component of the ground state, respectively.} The $\tilde{B}\,^2A_1$ state is described by Hund's case (b), and repumping lasers through this state address the $(K'=0,N'=0,J'=1/2,+)$ rotational level. The spin-rotation splitting in the ground state is 19~MHz \cite{CrozetXA2002}. All repumping lasers are tuned halfway between these two components and are sufficiently power-broadened to address both levels. The main cooling laser also consists of a single frequency component and can be scanned in the vicinity of one or both states. Hyperfine structure due to the nuclear spin of the hydrogen atoms is below the natural linewidth of the optical transitions and is therefore neglected \cite{NamikiHyperfine1998}.

Symmetry arguments within the $C_{3v}$ molecular symmetry group relate the nuclear spin state to the magnitude of $K_R$. Because rovibronic states with $K_R = 0$ transform like $A_1$ or $A_2$, and the total internal state, including nuclear spin, must also transform like $A_1$ or $A_2$ due to Fermi-Dirac statistics, the nuclear spin state corresponding to $K''=K_R=0$ is the ortho-CaOCH$_3$ NSI, which has nuclear spin $I=3/2$ and $A_1$ character. Similarly, rovibronic states with $|K_R|=1$ transform like $E$, meaning that the nuclear spin state must also have $E$ character. This corresponds to the para-CaOCH$_3$ NSI and total nuclear spin $I=1/2$~\cite{hougenDouble1980,Hirota,yu2019scalable}. Note that, because ortho~$\leftrightarrow$~para transitions are highly suppressed, nuclear spin statistics help to enforce the $\Delta K_R = 0$ selection rule.

We cool the para NSI by interacting with the $|K_R|=1$ stack. Because states with $K_R=\pm1$ come in unresolved opposite parity doublets, it is necessary to address more states to achieve rotational closure. Here we drive $^pP_{11}$, $^pQ_{12}$, and $^pP_{12}$ transitions from $K''=1$ to $K'=0$, addressing $N''=1$ and $N''=2$ in the ground state, as previously proposed in Ref. \cite{kozyryev2016proposal} (see Fig. 1 of main text). The spin-rotation splitting between ($N''=1,J''=1/2,3/2)$ in the ground state is 10~MHz and is bridged by a single power-broadened laser. The combined rotational line strength of the $^pP_{11}$ and $^pQ_{12}$ transitions from $N''=1$ is approximately 3x stronger than the $^pP_{12}$ transition from $N''=2$ \cite{brown1971rotational}.

\begin{figure}
\begin{centering}
\includegraphics[width = 0.35\textwidth]{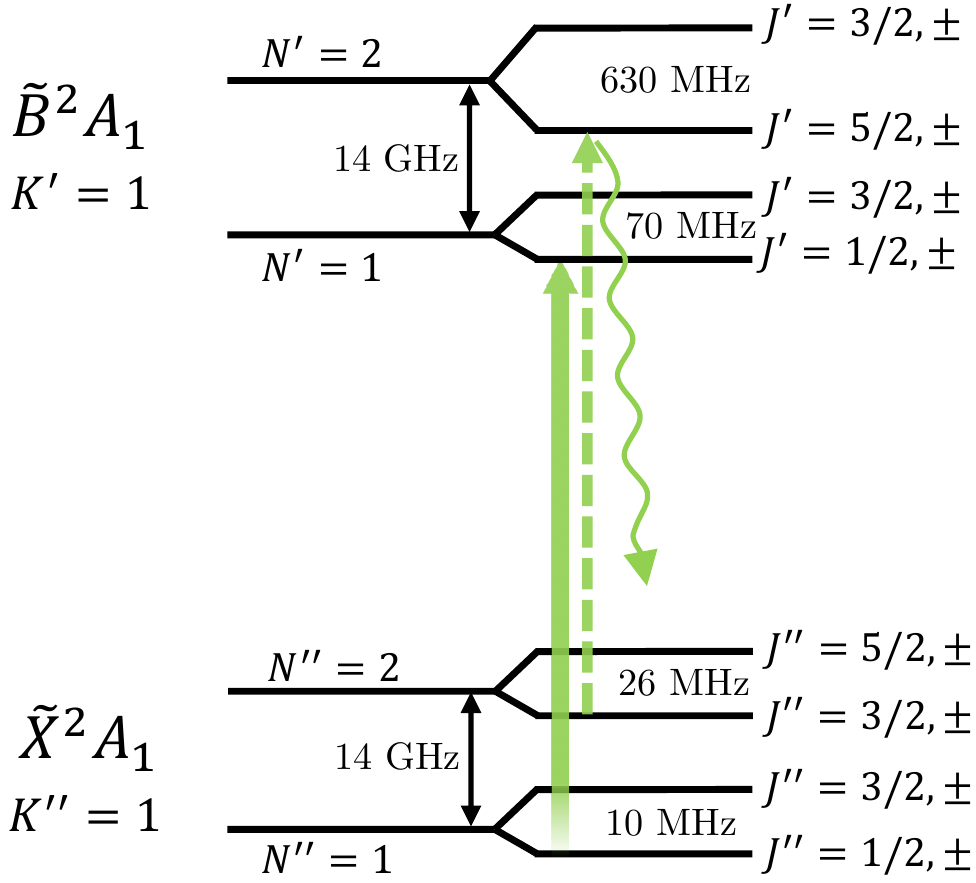}
\caption{\label{fig:Bstateproblem} \textbf{Limitations of $\mathbf{\textit{K}''=1}$ $\mathbf{\tilde{\textit{B}}}$ state cycling.} Due to a structural coincidence in CaOCH$_3$, the laser required to address $N''=1$ population (solid arrow) is near-resonant with another transition that pumps molecules from $N''=2$ to $J'=5/2$ and out of the cooling cycle (dashed arrow). This prevents laser cooling and many repumping pathways via the $\tilde{B}\,^2A_1$ state in the para-CaOCH$_3$ scheme used in this work.}
\par\end{centering}
\end{figure}

One repumping laser used in this work addresses the $\tilde{X}\,^2A_1(K''=1,N''=2,J''=3/2,\pm) \rightarrow \tilde{B}\,^2A_1(K'=1,N'=1,J'=1/2,\pm)$ transition through the $\tilde{B}\,^2A_1$ state. However, because the spin-rotation splitting between ($K'=1,N'=1,J'=1/2,3/2$) is only 70~MHz, excessive power broadening could lead to rotational loss via $J'=3/2$, and the intensity of this laser is therefore limited. Additionally, optical cycling via the $\tilde{B}\,^2A_1$ state for $|K_R|=1$ is limited in CaOCH$_3$ by the fact that the $N''=1 \rightarrow N'=1$ laser will also drive the nearby $N''=2 \rightarrow N'=2$ transition, tending to pump molecules out of the cooling cycle (Fig. \ref{fig:Bstateproblem}).

\begin{figure*}[ht]
\begin{centering}
\includegraphics[width = 0.8\textwidth]{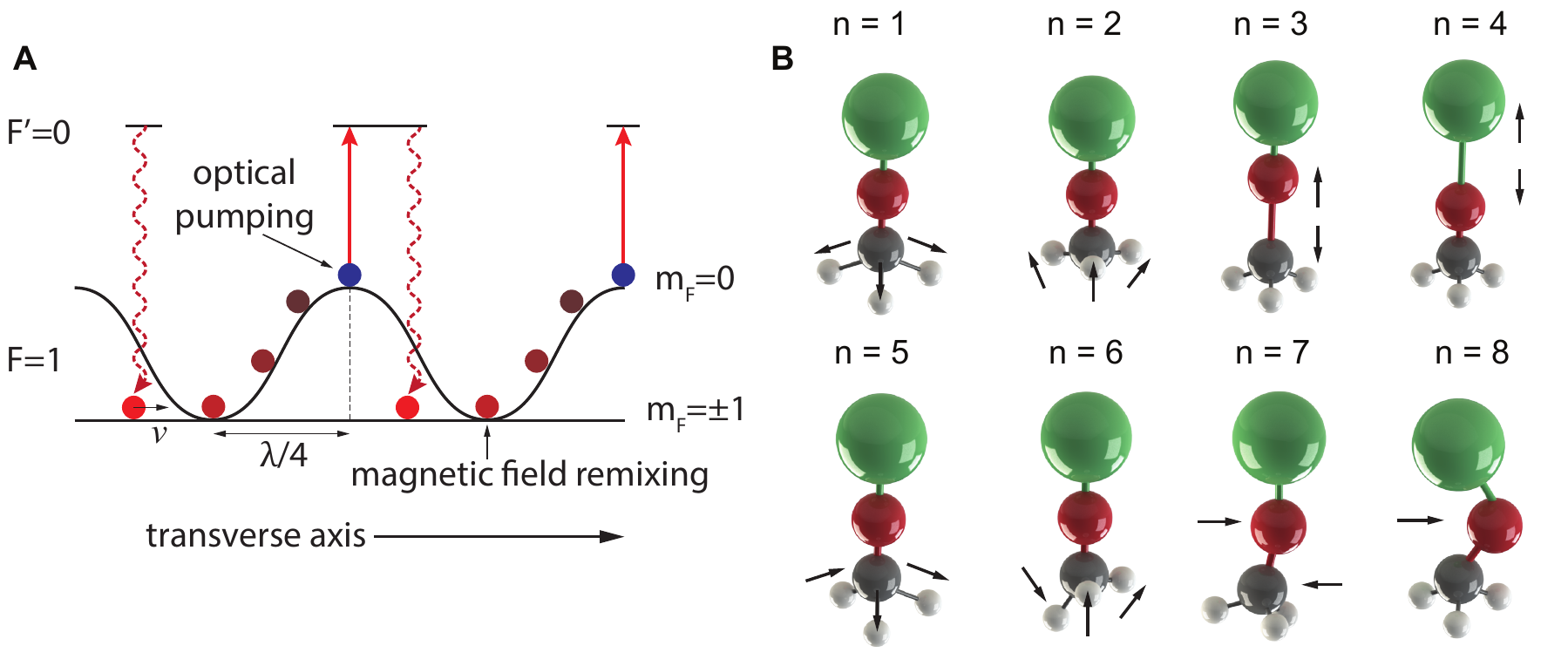}
\caption{\label{fig:scheme} \textbf{Sisyphus cooling scheme and CaOCH$_3$ vibrational normal modes.} \textbf{(A)} Sisyphus cooling scheme for the case of $\Delta>0$ for a $F=1 \rightarrow F'=0$ transition, as described in the text. \textbf{(B)} Normal modes of vibration for the CaOCH$_3$ molecule sorted by symmetry and in order of decreasing frequency. The first four modes are of $A_1$ symmetry, while the last four are of $E$ symmetry and are doubly degenerate, giving a total of 12 modes. Arrows represent the relative motion of specific atoms. The modes are enumerated as: $n=1$: CH$_3$ symmetric stretch, $n=2$: CH$_3$ symmetric bend, $n=3$: C-O (asymmetric) stretch, $n=4$: Ca-O (symmetric) stretch, $n=5$: CH$_3$ asymmetric stretch, $n=6$: CH$_3$ asymmetric bend, $n=7$: O-CH$_3$ wag, and $n=8$: Ca-O-C bend.}
\par\end{centering}
\end{figure*}

\bigskip
\noindent \textbf{CaOCH$_3$ vibrational modes}
\newline
To enumerate the vibrational modes of CaOCH$_3$, we follow the labeling convention given in Ref.  \cite{herzberg1966molecular}. In this convention, the vibrational modes are grouped by symmetry ($A_1$ and $E$) and sorted by decreasing frequency. CaOCH$_3$ has 12 ($=3N~-~6$, where $N$ is the number of atoms in the molecule) normal modes of vibration, four of which are of $A_1$ symmetry and four of which are doubly degenerate with $E$ symmetry. The modes have been sorted in order and depicted in Fig. \ref{fig:scheme}B. For this work, we only work with modes $n=3$, 4 and 8 as decays to other vibrational modes are highly suppressed due to symmetry and diagonal branching ratios.

\bigskip
\noindent \textbf{Experimental procedure} 
\newline
Details of the buffer gas beam source can be found in Ref. \cite{hutzler2012buffer}. We produce CaOCH$_3$ molecules by ablating a Ca metal target (Sigma-Aldrich, 99\%, natural isotopic abundance) with $\sim$~10~mJ of energy from a pulsed, second harmonic Nd:YAG laser. We flow $\sim$~6~standard cubic centimeters per minute (SCCM) of He buffer gas and $\sim$~0.01~SCCM of methanol (Sigma-Aldrich, $\geq$99.93\%) into a 50.8~mm long first-stage cryogenic buffer gas cell with 25.4~mm outer diameter. After thermalizing with He at $T\approx 2$~K in the cell, the resulting CaOCH$_3$ molecules are extracted through a 5~mm diameter aperture into a second-stage cell, where further collisions with lower density He buffer gas reduce the forward velocity to $150\pm 30$~m/s. The resulting cryogenic beam has a pulsed brightness of $\sim$$10^{10}$ molecules sr$^{-1}$state$^{-1}$s$^{-1}$. The molecule number in the beam is reduced to $\sim$10$^4$ after passing through a $3\times 3$~mm collimating aperture 35.5~cm from the cell. Molecules enter the cooling region 2~cm further downstream. Here, molecules experience a variable magnetic field produced by a circular coil with 15~cm diameter, mounted 10~cm from the cooling region and used to remix magnetic dark states. The standing wave used for Sisyphus cooling is composed of several frequencies as described in the main text. For ortho-CaOCH$_3$ cooling, the main cooling laser intensity is variable, while the intensity of each repumping laser is $\sim 400$~mW/cm$^2$. For para-CaOCH$_3$ each repumping laser has an intensity of $\sim 100$~mW/cm$^2$. The main cooling laser light is generated by the second harmonic of a 1258~nm Raman fiber amplifier (RFA) seeded by an external cavity diode laser (ECDL), while all repumping and imaging light is produced by continuous-wave (cw) dye lasers.

After cooling, the molecules propagate $\sim$10~cm before passing through a cleanup region containing multiple retroreflected passes of repumping light, spanning $\sim25$ mm in the longitudinal direction and with sufficient transverse width to address the full molecular beam. No magnetic field is applied, as the Earth's magnetic field is sufficient to remix dark states. The molecules are detected 41~cm further downstream by a circular, $\sim 11$~mm diameter beam with $\sim$ 20~mW of resonant light addressing the main cooling transition, $\tilde{X}\,^2A_1-\tilde{A}\,^2E_{1/2}0^0_0$, but produced using an independent laser. For para-CaOCH$_3$, this light contains $N''=1$ and $N''=2$ components, each with $\sim 8$ mW of power. To address all magnetic sublevels, the polarization of the light is switched between orthogonal linear polarizations at a rate of 1~MHz using a Pockels cell. In addition, we add $\sim$ 50-100~mW of $4_1^0$ repumping light to increase the number of scattered photons in the detection region (for para-CaOCH$_3$ this only contains the $N''=1$ component). Laser-induced fluorescence (LIF) photons are collected on an EMCCD camera (Andor iXon Ultra 897) for a duration of 15~ms starting 1~ms after the initial ablation pulse. The experiment is run at a repetition rate of $\sim 2$~s$^{-1}$, with shots alternating between Sisyphus cooled/heated and unperturbed configurations. To account for fluctuations in molecule numbers, beam images and traces depicted in Figs. 2-3 of the main text are the average of 150 successive shots, and each data point taken for parameter scans (Fig. \ref{fig:analysis} of the main text) is the average of 50 shots.

\bigskip
\noindent \textbf{Sisyphus mechanism and characteristic quantities}
\newline
The Sisyphus process for a simplified $F=1 \rightarrow F'=0$ transition is depicted in Fig. \ref{fig:scheme}A. The standing wave is linearly polarized and only addresses $\Delta m_F=0$ transitions, causing a sinusoidal, spatially varying AC Stark shift for molecules in the $m_F=0$ ground state. The $|m_F|=1$ states are dark. A molecule starting out at velocity $v$ in the bright state is forced to climb a potential hill (for $\Delta>0$), leading to a loss in kinetic energy. As it approaches the top of the hill, corresponding to an antinode of the standing wave, the molecule is optically pumped preferentially into a dark ground state, imparting the gained potential energy to the emitted photon. In the case of Sisyphus heating ($\Delta<0$), the opposite occurs and the molecule gains kinetic energy as it descends the hill before optical pumping occurs. This process is optimized if the time between scattering events $\gamma^{-1}$, where $\gamma$ is the scattering rate, corresponds to the travel time of the molecule from a node to an antinode of the potential, $\lambda/4v$. For ortho-CaOCH$_3$, this gives an estimated capture velocity $v_c = \lambda\gamma/4\approx$ 2~m/s, using the measured scattering rate of $\gamma/(2\pi) \approx 2\times10^6$~s$^{-1}$. This estimate is consistent with values reported in the main text.

Molecules undergo Larmor precession near nodes of the standing wave due to the external magnetic field, allowing remixing of the $m_F$ sublevels. This effect is optimized when the precession time, $(g_\text{eff}\mu_BB/\hbar)^{-1}$, is equal to the travel time of the molecule from antinode to node of the standing wave. Here $\mu_B$ is the Bohr magneton, $B$ is the applied magnetic field, $\hbar$ is the Planck constant, and $g_\text{eff} = 1/3$ is the degeneracy-weighted Land\'e $g$-factor of the ground state hyperfine components of CaOCH$_3$. This gives an estimate for the optimal B field $B_0 = 4\hbar v/(g_\text{eff}\mu_B\lambda) \approx 1.6$~G for ortho-CaOCH$_3$, assuming the average velocity is $v_c/2$ \cite{lim2018laser}.

We note that other sub-Doppler processes like grey molasses and polarization gradient cooling can also occur in our system \cite{dalibard1989}. However, for the linearly polarized intensity standing wave and finite magnetic fields used in this work, the magnetically assisted Sisyphus effect dominates.

\bigskip
\noindent \textbf{Beam fitting procedure}
\newline
Analyzed beam widths are determined by fitting Gaussian distributions to the 1D beam profiles obtained by integrating out the longitudinal dimension of the EMCCD beam images. The unperturbed beam profiles are fit to a single Gaussian distribution to obtain their width $\sigma_\text{unp}$. The traces pertaining to Sisyphus heating or cooling are instead fit to a sum of Gaussian distributions.

For $\Delta > 0$, there is a narrow central feature (width $\sigma_n$) corresponding to the Sisyphus-cooled molecules within $v_c$ and a broader feature (width $\sigma_w$) representing the uncooled molecules. Near the center of the cloud, the number of uncaptured molecules goes smoothly to zero because all low-velocity molecules are captured by the Sisyphus cooling. We represent this feature as a dip in the spatial distribution of the uncaptured molecules, which goes to zero at the spatial center of the beam with a characteristic width equal to that of the Sisyphus-cooled molecules. To capture these effects, we fit the beam profiles to the expression
\begin{align}
y = &y_0 \nonumber \\
&+ \underbrace{A\left(\text{exp}[-(x-x_c)^2/2\sigma_w^2] - \text{exp}[-(x-x_c)^2/2\sigma_n^2]\right)}_\text{uncaptured molecules} \nonumber \\
&+ \underbrace{B\text{exp}[-(x-x_c)^2/2\sigma_n^2]}_\text{captured molecules},
\label{eq:gaussian2}
\end{align}
where $y_0$ is an offset, $A$ and $B$ are fit amplitudes, and $x_c$ is the center position. Treating the Sisyphus cooled and uncaptured molecules separately in this way allows us to define a cooled fraction for the case of $\Delta>0$. This is the ratio of the area under the narrow Gaussian (corresponding to captured molecules) to the total area (corresponding to all molecules). This quantity has physical meaning and depends on the capture velocity $v_c$ and the efficiency of the cooling process.

\begin{figure}
\begin{centering}
\includegraphics[width = 1\columnwidth]{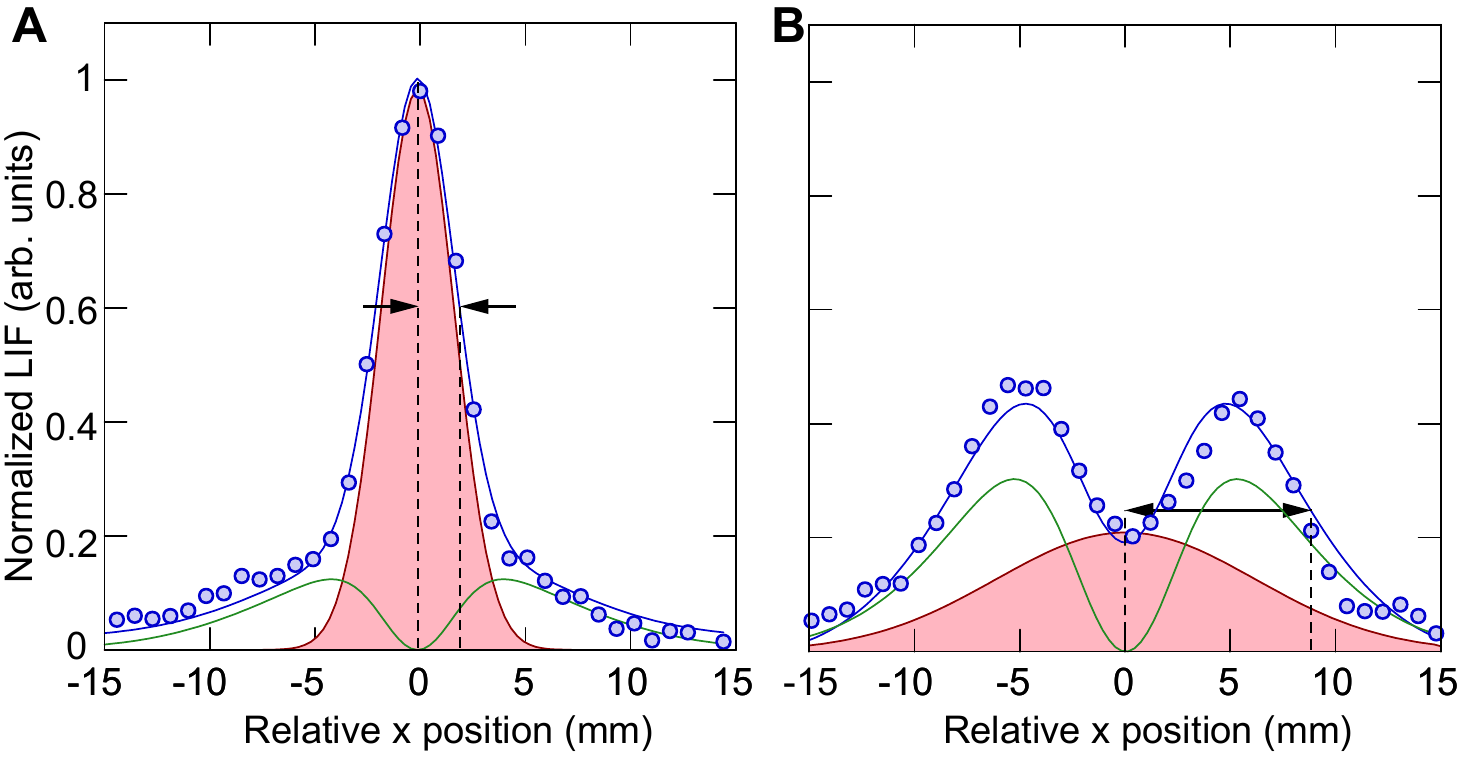}
\caption{\label{fig:fitting} Curve fitting protocol. The fitting procedure is illustrated for \textbf{(A)} a Sisyphus cooled beam profile at $\Delta=+25$~MHz and \textbf{(B)} a Sisyphus heated profile at $\Delta=-15$~MHz. Plotted are the data (circles), overall fit (blue curve), uncaptured contribution (green curve), and captured contribution (pink shaded region). The vertical axis for both panels is normalized to the Sisyphus cooled peak. For Sisyphus cooling ($\Delta>0$) only, the cooled fraction is defined as the shaded area divided by the total area. In addition, arrows define the model-independent beam width. For cooling \textbf{(A)}, this is the distance from the peak at which the normalized LIF signal drops to $1/\sqrt{e} \approx 0.61$. For heating \textbf{(B)}, this is the furthest distance from the center at which the LIF signal drops to $1/\sqrt{e}$ of the mean height of the two spatially separated peaks.}
\par\end{centering}
\end{figure}

This functional form also captures the bimodal structure of the Sisyphus heated ($\Delta < 0$) traces well, although the physical interpretation of captured vs. uncaptured molecules is less precise in this case. Physically, the molecules within $v_c$ are heated outwards and start accumulating at a point where Doppler cooling balances Sisyphus heating. Fits in both cases are illustrated in Fig. \ref{fig:fitting}.

In order to have a consistent definition of beam width when comparing Sisyphus heated and cooled configurations, we define the width as the $1/\sqrt{e}$ radius of the beam profile. For the case of cooling, this is defined relative to the height of the large central peak, while for heating it is defined with respect to the average peak height of the two side lobes (Fig. \ref{fig:fitting}). Both the $1/\sqrt{e}$ beam width and the cooled fraction are used for comparison to the optical Bloch equation simulation (see Fig. \ref{fig:analysis} of the main text and the discussion below).

The resonance condition ($\Delta=0$) is determined by matching the perturbed beam width to the unperturbed beam (grey shaded region in Fig. \ref{fig:analysis}A of the main text). We determine this frequency to be 5~MHz blue-detuned from the $\tilde{X}\,^2A_1 0_0 (J=3/2)$ state and 15~MHz red-detuned from the $\tilde{X}\,^2A_1 0_0 (J=1/2)$ component, which is consistent with the former state having twice as many sublevels.

\begin{figure}
\begin{centering}
\includegraphics[width = 1\columnwidth]{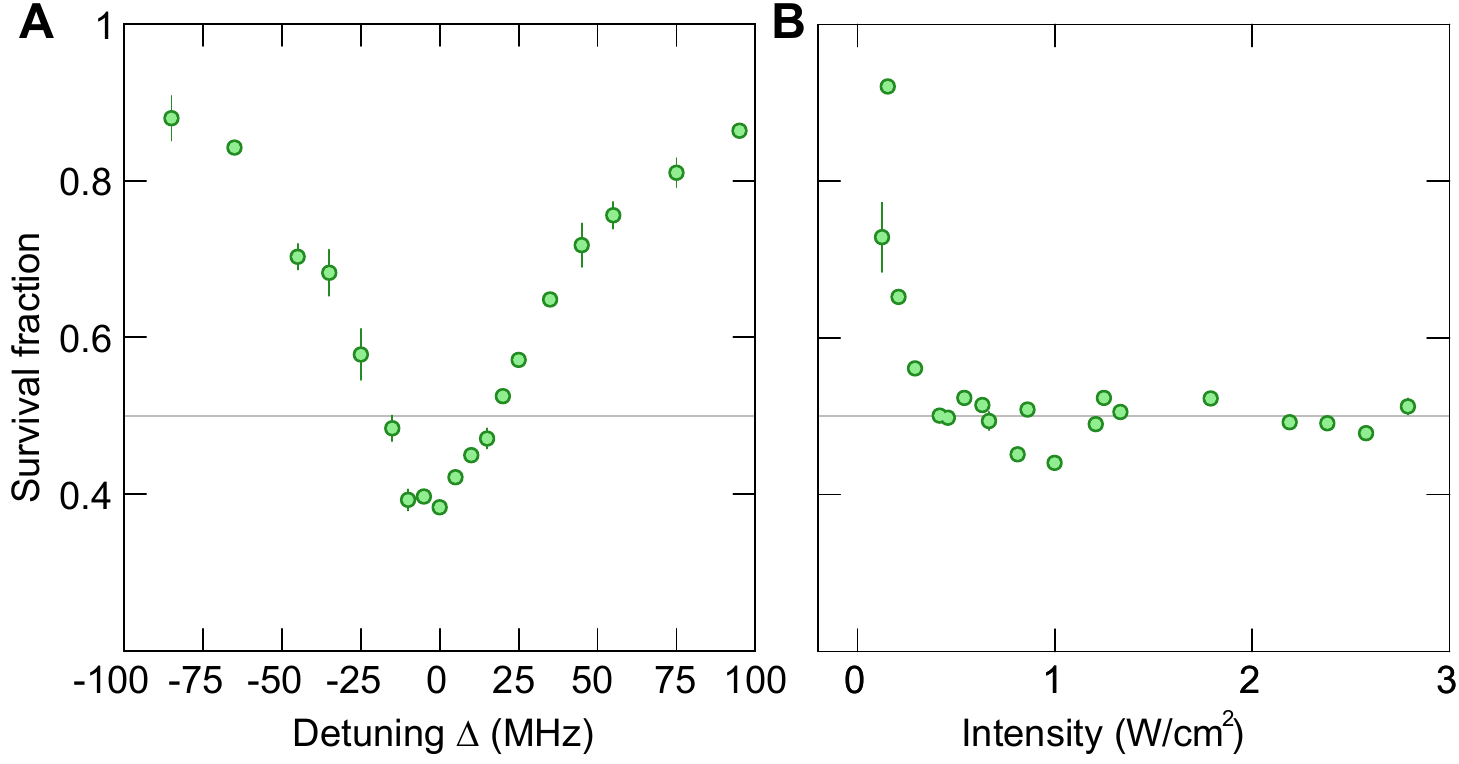}
\caption{\label{fig:repfrac} Variation of survival fraction for ortho-CaOCH$\mathbf{_3}$. \textbf{(A)} Variation of the survival fraction with detuning $\Delta$ at an intensity of 400~mW/cm$^2$ and a remixing field of $|B|=1$~G. On resonance, the survival fraction reaches a minimum of 38(1)\%. It increases away from resonance as the scattering rate decreases. \textbf{(B)} Variation with cooling laser intensity at $\Delta=+25$~MHz and $|B|=1$~G. The survival fraction saturates to 49(3)$\%$ for intensities greater than $\sim$~600~mW/cm$^2$. In both panels, the grey line indicates a survival fraction of 50$\%$.}
\par\end{centering}
\end{figure}

\bigskip
\noindent \textbf{Estimation of photon number}
\newline
We estimate the number of scattered photons from the survival fraction, defined as the ratio of the integrated area of the Sisyphus cooled/heated images to that of the unperturbed beam. This provides a measure of the number of molecules remaining in the optical cycling transition, or equivalently, the fraction of molecules lost to vibrational dark states. The survival fraction is directly related to the number of scattered photons via vibrational branching ratios (VBRs), and its variation with detuning and intensity is shown in Fig. \ref{fig:repfrac}. For optimized Sisyphus cooling parameters of 600~mW/cm$^2$, $\Delta=+25$~MHz, and $|B| = 1$~G, we measure a survival fraction of 49(3)$\%$, while the survival fraction reaches a minimum value of 38(1)$\%$ on resonance where the scattering rate is maximized.

We model photon scattering as a Bernoulli process starting with the VBRs measured in Ref. \cite{kozyryev2019determination}. In order to estimate the effect of small decay channels not observed in that work, we include an additional branching ratio $r_\text{other}$ characterizing loss to all such states. We then vary this parameter subject to two conditions: (1) the remaining VBRs, when properly scaled to ignore the effect of $r_\text{other}$, agree within error with Ref. \cite{kozyryev2019determination}, and (2) the results agree with measurements taken over the course of this work characterizing the loss to excited vibrational levels after photon cycling on some subset of known states. Specifically, we measure that after depleting all population from the $\tilde{X}\,^2A_1 0_0$ state via photon cycling on the $\tilde{X}\,^2A_1-\tilde{A}\,^2E_{1/2} 0^0_0$ vibronic transition, 80(10)\% of molecules are found in the $\tilde{X}\,^2A_1 4_1$ state. Additionally, after cycling photons on both the $\tilde{X}\,^2A_1-\tilde{A}\,^2E_{1/2} 0^0_0$ main and $\tilde{X}\,^2A_1-\tilde{B}\,^2A_1 4^0_1$ repumping transitions, 50(10)\% of lost molecules can be recovered from the $\tilde{X}\,^2A_1 3_1$ state. Using this procedure, we infer VBRs for decay from the $\tilde{A}\,^2E_{1/2}$ state, given in Table \ref{tab:vbrs}. We cannot independently determine the VBRs from the $\tilde{B}\,^2A_1$ state and fix these at the values from Ref. \cite{kozyryev2019determination} while conservatively including an additional loss $r_\text{other}^B = 0.005$, though this only weakly affects the final result.

It is estimated from harmonic VBR calculations similar to those used in Ref. \cite{kozyryev2019determination} that 4 additional repumping lasers are required to scatter 1000 photons on average from CaOCH$_3$. These address the $3_1 4_1 8_1$, $4_2$, $7_2$, and $2_1$ vibrational modes of the $\tilde{X} \,^2A_1$ state.

\begin{table}
\centering
\begin{tabular}{| c | c |}
\hline
Decay & Branching Ratio \\
\hline
$0_0^0 \tilde{A}^2E_{1/2} \rightarrow \tilde{X}^2A_1$ & $0.925(7)$ \\
$4_1^0 \tilde{A}^2E_{1/2} \rightarrow \tilde{X}^2A_1$ & $0.060(2)$ \\
$3_1^0 \tilde{A}^2E_{1/2} \rightarrow \tilde{X}^2A_1$ & $8.6^{+0.7}_{-3.2} \times 10^{-3}$ \\
Other & $6.4^{+5.3}_{-5.0} \times 10^{-3}$ \\
\hline
\end{tabular}
\caption{Approximate branching ratios determined for CaOCH$_3$ laser cooling transitions as described in the text. These rely on both existing dispersed fluorescence measurements \cite{kozyryev2019determination} as well as independent optical pumping measurements conducted over the course of the present work.}
\label{tab:vbrs}
\end{table} 

\bigskip
\noindent \textbf{Effective scattering rates} \\
The maximum scattering rate in a multi-level system can be related to the total number of ground states $n_g$ and excited states $n_e$ by the expression $\Gamma_\text{max} = \Gamma \frac{n_e}{n_g+n_e}$, where $\Gamma$ is the natural linewidth of the main cycling transition \cite{norrgard2016submillikelvin}. In the ortho-CaOCH$_3$ ($K''=0$) cooling scheme, including hyperfine structure with $I=3/2$ we have $n_g = 24$ and $n_e = 8$, leading to a maximum scattering rate $\Gamma_\text{max}^{(0)} = \Gamma/4$. In the para-CaOCH$_3$ case with $I=1/2$, we couple the $N''=1(J''=1/2,3/2)$ and $N''=2(J''=3/2)$ manifolds of the $\tilde{X}\,^2A_1 0_0$ ground state, as well as the $N''=1(J''=1/2,3/2)$ components of the $\tilde{X}\,^2A_14_1$ level, to the $\tilde{A}\,^2E_{1/2}$ state. After accounting for the parity doubling of every level, this gives $n_g=64$ and $n_e=8$; therefore $\Gamma_\text{max}^{(1)} = \Gamma/9$. Assuming $\Gamma$ is the same for both cycling schemes, we therefore expect a reduction of $\Gamma_\text{max}^{(0)} / \Gamma_\text{max}^{(1)} \approx 9/4$ in scattering rate for para-CaOCH$_3$ cycling compared to the ortho NSI.

\bigskip
\noindent \textbf{Temperature determination} \\
In order to determine the temperature of the molecules, we perform a Monte Carlo (MC) simulation of the molecular beam propagation in 3D and match the transverse width in the detection region to our data. We then use the velocity spread of the MC simulated beam, $\sigma_v = \sqrt{k_B T_\perp/m}$, where $k_B$ is the Boltzmann constant and $m$ is the mass of CaOCH$_3$, to determine the beam temperature $T_\perp$. After initial beam propagation through the $3\times3$ mm collimating aperture, we find a spatial beam width $\sigma_x = 6.1(1)$~mm, corresponding to a temperature $T_\perp = 22\pm 1$ mK.

We use two separate methods to determine the temperature of the cooled beam; differences between the two reflect systematic errors associated with these results. In one approach, the transverse velocity of each molecule is divided by a constant factor (chosen to match the detected $1/\sqrt{e}$ beam width with the experimental data) as it passes through the cooling region, reducing the temperature but preserving correlations between the position and velocity of the molecules. This gives a temperature of $T_\perp = 1.20(35)$ mK, where the reported error is statistical only. In the other approach, molecules are randomly assigned velocities from a thermal distribution with temperature $T_\perp$ when they enter the cooling region, with $T_\perp$ chosen to match the observed $1/\sqrt{e}$ width in the detection region. This approach entirely scrambles correlations between position and velocity which may exist after cooling, and gives a temperature $T_\perp = 2.4(5)$ mK. Averaging these results gives a final temperature estimate $T_\perp = 1.8 \pm 0.7$ mK. This agrees well with the final temperature predicted from the optical Bloch equation simulations described below and in the main text.

\bigskip
\noindent \textbf{Simulation} \\

We model ortho-CaOCH$_3$ laser cooling by solving optical Bloch equations for the density matrix describing the internal state evolution of a molecule moving through the laser cooling region. Further details of this method may be found in Refs. \cite{emile1993,Devlin2016,Devlin2018}. Because power broadening is significant compared to the 19 MHz separation between the $J''=3/2$ and $J''=1/2$ ground states, we compute force profiles approximating the system as a single $J'' = 3/2 \leftrightarrow J' = 1/2$ transition, which captures the essential physics of Sisyphus laser cooling due to magnetic dark states in the system. The average force and scattering rate is computed once the molecular operators reach a periodic steady state that tracks the periodicity of the molecule-light Hamiltonian. This procedure is repeated for the full transverse velocity distribution to obtain force profiles used to simulate the propagation of molecules through the cooling light. The simulation does not implicitly take into account the reduced scattering rate of a multi-level system, so the force profiles are instead explicitly scaled according to the effective scattering rate $\Gamma^{(0)}_\text{max} = \Gamma/4$, as discussed above. This results in good agreement between the simulated and experimentally measured scattering rates. Similarly, we used an effective saturation intensity $I_\text{sat,eff} = 1.2 I_\text{sat, 2-level}$ in the simulation, which is close to the multi-level correction factor $2n_g/(n_g + n_e) = 3/2$ expected from a simple rate equation analysis \cite{norrgard2016submillikelvin,Baum2020}. Finally, we found that an effective Land\'e $g$-factor of $g_J = 0.85$ agreed best with the data. This is somewhat larger than the low-field value of $g_J = 2/3$, but may be explained by noting that small magnetic fields of a few Gauss are sufficient to begin decoupling the electron spin due to the small spin-rotation interaction energy in CaOCH$_3$. Hyperfine interactions are unresolved across the experimental parameters investigated and are therefore not included in the model. 

The experimental setup is modeled by a Monte Carlo simulation of the full three-dimensional CaOCH$_3$ molecular beam. The beam is initialized at the cell exit aperture and ballistically propagated through the collimating aperture to the laser cooling region. Any molecules that do not successfully pass through the aperture are discarded. Interaction with the cooling light is modeled using the optical Bloch equations described above. Once the molecules have left the cooling region, they ballistically propagate to the detection region, where smoothed histograms are computed to reproduce spatial beam profiles similar to those of Fig. 3 in the main text. Simulated beam widths and cooled fractions are computed for the smoothed profiles in the same manner as for the experimental data. The lower and upper bounds for the shaded areas shown in Fig. \ref{fig:analysis} are computed by scaling the average force profile by the relative uncertainty in the number of scattered photons quoted in the main text ($\sim 0.6$ and $\sim 2.3$ for the lower and upper bounds, respectively). For all calculations, a constant offset field of 0.4~G was assumed in addition to the variable external field to account for Earth's magnetic field, especially important around $B=0$.

From the simulations, we find that the damping coefficient at optimal detuning and magnetic field is $\eta_\text{max} \approx 3 \times 10^5 \text{ s}^{-1}$, which is several orders of magnitude higher than those achieved for standard radiative cooling of linear molecules, but expected for the Sisyphus mechanism \cite{truppe2017molecules,anderegg2017radio,Baum2020,emile1993}. From the force profiles, we also infer a capture velocity $v_c \sim 1.9$ m/s, which is consistent with the experimental estimate obtained from the beam profiles for the case of  $\Delta < 0$.

\end{document}